\documentclass[preprint, 12pt,  authoryear]{elsarticle}

\makeatletter
\def\ps@pprintTitle{%
 \let\@oddhead\@empty
 \let\@evenhead\@empty
 \def\@oddfoot{}%
 \let\@evenfoot\@oddfoot}
\makeatother

\usepackage{amsmath, amssymb, amsfonts, amsthm, mathtools, bm}
\usepackage{centernot}
\usepackage[thinc]{esdiff}
\usepackage{bbm} % instead of mathbbol (Overleaf-safe)

\usepackage{graphicx}
\usepackage{graphics}
\usepackage{tikz}
\usetikzlibrary{arrows,calc,shapes.geometric,positioning,trees,shadows,fit,arrows.meta,decorations}
\usepackage{forest}
\usepackage{subcaption}   % modern, don't combine with subfig
\usepackage{caption}
\usepackage{adjustbox}
\usepackage{pdflscape}
\usepackage{rotating}
\usepackage{float}

\usepackage{array}
\usepackage{longtable}
\usepackage{multirow}
\usepackage{booktabs}
\usepackage[flushleft]{threeparttable}
\usepackage{tablefootnote}

\usepackage[margin=1in]{geometry}
\usepackage{setspace}
\usepackage{titlesec}
\titlespacing*{\section}{0pt}{*1}{*0}
\titlespacing*{\subsection}{0pt}{*1}{*0}

\usepackage{afterpage}
\usepackage{changepage}
\usepackage{ulem}
\usepackage{tcolorbox}
\usepackage{color}
\usepackage{comment}
\usepackage{multibib}
\usepackage[toc, header]{appendix}

\usepackage[authoryear]{natbib}
\usepackage[hidelinks]{hyperref}

\def\targetedEstimand{\gamma_{t}}
\def\estimandOfInterest{\gamma_{i}}
\def\claimedEstimand{\gamma_{c}}
\def\claimedEstimandAbstract{\gamma_{c}^{a}}
\def\claimedEstimandResults{\gamma_{c}^{r}}
\def\claimedEstimandDiscussion{\gamma_{c}^{d}}
\def\LATE{\Psi_{LATE}}
\def\ATE{\Psi_{ATE}}

\def\ind{\perp \!\!\! \perp}
\def\notInd{\not\!\perp\!\!\!\perp}

\begin{document}

\begin{frontmatter}
\title{Interpretational errors with instrumental variables}

\author[1]{Luca Locher}
\author[2]{Mats Stensrud}
\author[3]{Aaron Sarvet}

\address[1]{Epidemiology, Biostatistics and Prevention Institute, University of Zurich, Switzerland}
\address[2]{Institute of Mathematics, École Polytechnique Fédérale de Lausanne, Switzerland}
\address[3]{Department of Biostatistics \& Epidemiology, University of Massachusetts, Amherst, MA}

\begin{abstract}

Instrumental variables (IV) are often used to identify causal effects in observational settings and experiments subject to non-compliance. Under canonical assumptions, IVs allow us to identify a so-called local average treatment effect (LATE). The use of IVs is often accompanied by a pragmatic decision to abandon the identification of the causal parameter that corresponds to the original research question and target the LATE instead. This pragmatic decision presents a potential source of error: an investigator mistakenly interprets findings as if they had made inference on their original causal parameter of interest. We conducted a systematic review and meta-analysis of patterns of pragmatism and interpretational errors in the applied IV literature published in leading journals of economics, political science, epidemiology, and clinical medicine (n = 309 unique studies). We found that a large fraction of studies targeted the LATE, although specific interest in this parameter was rare. Of these studies, 61\% contained claims that mistakenly suggested that another parameter was targeted—one whose value likely differs, and could even have the opposite sign, from the parameter actually estimated. Our findings suggest that the validity of conclusions drawn from IV applications is often compromised by interpretational errors.
\end{abstract}
\end{frontmatter}

\doublespacing
\section{Introduction}

Strategies for identifying causal effects from experimental and quasi-experimental data have undoubtedly enriched the applied statistician's toolbox and their popularization has likely increased the internal validity of empirical investigations. Many methods associated with this “credibility revolution” \citep{angrist_credibility_2010} draw attention to causal estimands that arise as side effects of the identification strategy, not from deliberate choice. One such estimand is the local (L) average treatment effect (ATE), a parameter commonly targeted in instrumental variable (IV) analyses. \cite{imbens_causality_2022} summarizes the pragmatic rationale behind the LATE: \textit{``The standard approach to identification was to first state what the target estimand was, and then to articulate the identification strategy through assumptions that would allow one to identify that estimand. Angrist and I turned this strategy around and introduced an alternative way to study identification questions. Rather than start with an estimand and ask if and how we could identify that, we started by asking what we could identify under reasonable assumptions.''}

Pragmatism is a vital component in the application of statistics, distinguishes it from pure mathematical disciplines, and reinforces its status as a methodological discipline in close alliance with many applied domains. However, one line of criticism raises the concern that in the case of LATE estimation, the pragmatic pursuit of identification compromises relevance—yielding an identified quantity of diminished informative value (\cite{deaton2010instruments}, \cite{heckman_comparing_2010}, \cite{swanson_challenging_2018}). Here, we focus on a different, albeit related, problem arising from distinctly pragmatic approaches to causal inference. Suppose an investigator is interested in a causal parameter, say the ATE. Concerned that the ATE may not be identified under plausible assumptions, the investigator instead targets the LATE. We argue that the shift to the LATE presents a source of interpretational error: the investigator may interpret findings as if they pertain to the original parameter of interest. We refer to this type of error as \textit{identity slippage}.

In this article, we study patterns of \textit{pragmatism} and \textit{identity slippage}, and present empirical evidence that both phenomena are widespread in the applied literature. We also discuss the practical consequences of these phenomena. 

Our investigation is structured  as follows. First, we revisit assumptions for the identification of causal effects with IV. Then, we proceed to address three central questions: (1) \textit{Why is identity slippage relevant?} It may create the false impression that inference pertains to another parameter—one with a value different from the parameter under analysis and therefore corresponding to different policy questions. Thus, \textit{identity slippage} poses a risk to the translation of scientific findings into actionable implications, with perhaps the same impact as other, more well-known errors arising from, for example, model misspecification or p-hacking. (2) \textit{How frequently does it occur in applied research?} We studied an extensive sample of IV applications published in leading journals of disciplines where IV applications are popular (economics, epidemiology, medicine, and political science). We found that \textit{identity slippage} was common across all of these domains; 61\% of the studied articles exhibited such instances. (3) \textit{What factors contribute to its prevalence?} We identify \textit{pragmatism} as a cause of \textit{identity slippage} and argue that the subtle interpretation of the LATE might increase the chance of making such an error.

\section{Identification of causal effects with IV}
\label{identification_iv}

IVs are a popular tool for the identification of causal effects in observational settings and in experiments subject to non-compliance. In this section, we revisit assumptions for the identification of causal effects with IVs, focusing specifically on the (L)ATE. Although IV methods can, in principle, be used to target other estimands, such as marginal treatment effects (MTEs) \citep{heckman_chapter_2007}, these are beyond the scope of this article.

We use capital letters to denote random variables and lower-case letters to denote their possible realizations. We consider an instrument $Z$, a treatment $A$ and an outcome $Y$. We use parentheses to denote potential outcome variables. For example, $Y(z, a)$ denotes the potential outcome under an intervention that sets $Z=z$ and $A=a$, and $A(z)$ denotes the potential treatment under an intervention that sets $Z = z$. \cite{angrist_identification_1996} show that a valid IV allows for identification of a so called LATE, here denoted as $\LATE$. According to  \citet{angrist_identification_1996}, an IV is valid if the following conditions hold:
\begin{align}
    &\text{(Stable unit treatment value)} \label{ass:1} \\ 
    &\quad \text{(a)} \quad \text{If } Z = z, \text{ then } A = A(z), \text{ for all }  z, \text{ with probability 1},\nonumber \\
    &\quad \text{(b)} \quad \text{If } Z = z \text{ and } A = a, \text{ then } Y = Y(z, a), \text{ for all }  z, a, \text{ with probability 1},\nonumber \\
    & \text{(Relevance)} \quad Z \notInd A, \\
    & \text{(Independence)} \quad Z \ind Y(a), \\
    & \text{(Exclusion restriction)} \quad Y(z, a) = Y(a) \text{ for all }  z, a, \text{ with probability 1},\\
    & \text{(Monotonicity)} \quad A(z) \geq A(z') \text{ for all } z \geq z', \text{ with probability 1}. \label{ass:5}
\end{align}

In a setting with binary $Z$ and $A$, where Z is a causal instrument, i.e. $Z$ and $A$ are not merely associated via a non-causal path, $\LATE$ can be expressed as: 
\begin{align}
\LATE = E[Y(1) - Y(0) | A(1) > A(0)].
\end{align}

The interpretation of this quantity is relatively straightforward: It corresponds to the ATE in the sub-population of so-called compliers, i.e. the units that are exposed to $A$ if and only if they are exposed to $Z$ \citep{angrist_identification_1996}. The population of compliers is in general not identifiable, but the size and observed characteristics of this subgroup can be estimated \citep{swanson_challenging_2018}.

In a generalized setting where either $A$ or $Z$ are multi-valued or a set of control covariates is introduced, the targeted quantity represents a non-negatively weighted average of local effects and the interpretation becomes more nuanced. \cite{angrist_mostly_2009} provide a comprehensive overview of the interpretation of $\LATE$ in these settings. Further complications are added if instruments are non-causal \citep{swanson_challenging_2018}\footnote{As an example of a non-causal instrument, consider judge-leniency designs. The quantity used as instrument, e.g. a judge's denial rate in other cases is not a direct cause of the studied treatment. Rather, the treatment and the instrument share a common cause, the latent `leniency' of the judge}. Recent work by \cite{blandhol_when_2022} demonstrates that in settings involving control covariates, the identified quantity retains its interpretation as a non-negative average of local effects only when covariates are controlled for non-parametrically.

Alternatively, IV can be used to target the ATE, here denoted as $\ATE$. In settings with discrete data, partial identification approaches can be used to calculate sharp bounds of $\ATE$ with no additional assumptions (\cite{manski_nonparametric_1990}, \cite{swanson2018partial}, \cite{duarte_automated_2024}). Point identification of $\ATE$ requires additional assumptions that restrict the heterogeneity of treatment effects. While constant linear effects are a standard example, weaker assumptions that allow for point identification exist, such as those presented in \cite{wang_bounded_2018}.

\section{Pragmatism in causal inference and identity slippage}
\label{identity_slippage}

As demonstrated in the previous section, an IV can be used to target both $\LATE$ and $\ATE$. Consider now an investigator who is interested in $\ATE$, but decides that the assumptions required for point identification of $\ATE$ are implausible. In what follows, we distinguish between two types of studies, based on the approach adopted in such a context: those that follow a \textit{pragmatic approach of causal inference}, and those that adhere to a \textit{strict approach of causal inference}. \textit{Pragmatic} studies respond by shifting their focus to $\LATE$. In contrast, \textit{strict} studies retain their focus on $\ATE$, and either opt for partial identification, force point identification by stating implausible assumptions, or abandon the research objective altogether. The \textit{pragmatic} approach aligns with the perspective expressed by \cite{imbens_causality_2022}, as cited in the introduction, while the \textit{strict} approach reflects the principle articulated by \cite{heckman_comparing_2010}, who argue: “The primary question regarding the choice of an empirical approach to analyzing economic data should be \textit{What economic question does the analyst seek to answer?}”

\begin{figure}[t]
    \footnotesize
    % First subfigure (a)
    \begin{subfigure}[t]{0.22\textwidth}
        \begin{tikzpicture}
            \node at (0,4.8) {Research objective};
            \node at (0,2.7) {Statistical analysis};
            \node at (0,0.6) {Interpretation};
            \node at (0,0) {};
        \end{tikzpicture}
    \end{subfigure}
    \begin{subfigure}[t]{0.2\textwidth}
        \centering
        \begin{tikzpicture}[
        inactive/.style={circle, draw, fill=gray!20, text=darkgray, minimum size=1.5cm, text centered},
        active/.style={circle, draw, minimum size=1.5cm, text centered},
        node distance=0.6cm and 0.15cm
        ]
        % Left panel
        \node[active] at (0,0)  (ATE1) {$\ATE$};
        \node[inactive, right=of ATE1] (LATE1) {$\LATE$};
        
        \node[inactive, below=of LATE1] (LATE2) {$\LATE$};
        \node[active, below=of ATE1] (ATE2) {$\ATE$};
        
        \node[inactive, below=of LATE2] (LATE3) {$\LATE$};
        \node[active, below=of ATE2] (ATE3) {$\ATE$};
        
        \draw[->, thick] (ATE1) -- (ATE2);
        \draw[->, thick] (ATE2) -- (ATE3);

        \end{tikzpicture}
        \caption{}
        \label{fig:levels_a}
    \end{subfigure}
    \hfill
    \begin{subfigure}[t]{0.2\textwidth}
        \centering
        \begin{tikzpicture}[
        inactive/.style={circle, draw, fill=gray!20, text=darkgray, minimum size=1.5cm, text centered},
        active/.style={circle, draw, minimum size=1.5cm, text centered},
        node distance=0.6cm and 0.15cm
        ]
        % Left panel
        \node[active] at (0,0)  (ATE1) {$\ATE$};
        \node[inactive, right=of ATE1] (LATE1) {$\LATE$};
        
        \node[active, below=of LATE1] (LATE2) {$\LATE$};
        \node[inactive, below=of ATE1] (ATE2) {$\ATE$};
        
        \node[active, below=of LATE2] (LATE3) {$\LATE$};
        \node[inactive, below=of ATE2] (ATE3) {$\ATE$};
        
        \draw[->, thick] (ATE1) -- (LATE2);
        \draw[->, thick] (LATE2) -- (LATE3);

        \end{tikzpicture}
        \caption{}
        \label{fig:levels_b}
    \end{subfigure}
    \hfill
    \begin{subfigure}[t]{0.2\textwidth}
        \centering
        \begin{tikzpicture}[
        inactive/.style={circle, draw, fill=gray!20, text=darkgray, minimum size=1.5cm, text centered},
        active/.style={circle, draw, minimum size=1.5cm, text centered},
        node distance=0.6cm and 0.15cm
        ]
        % Left panel
        \node[active] at (0,0)  (ATE1) {$\ATE$};
        \node[inactive, right=of ATE1] (LATE1) {$\LATE$};
        
        \node[active, below=of LATE1] (LATE2) {$\LATE$};
        \node[inactive, below=of ATE1] (ATE2) {$\ATE$};
        
        \node[inactive, below=of LATE2] (LATE3) {$\LATE$};
        \node[active, below=of ATE2] (ATE3) {$\ATE$};
        
        \draw[->, thick] (ATE1) -- (LATE2);
        \draw[->, thick] (LATE2) -- (ATE3);

        % Labels for the rows
        \end{tikzpicture}
        \caption{}
        \label{fig:levels_c}
    \end{subfigure}
    \hfill
    \caption{Illustration of a \textit{pragmatic approach} and \textit{identity slippage}. A pragmatic approach is characterized by a discordance between research objective and statistical analysis, and \textit{identity slippage} is indicated by a discordance between statistical analysis and result interpretation.}
    \label{fig:levels}
\end{figure}
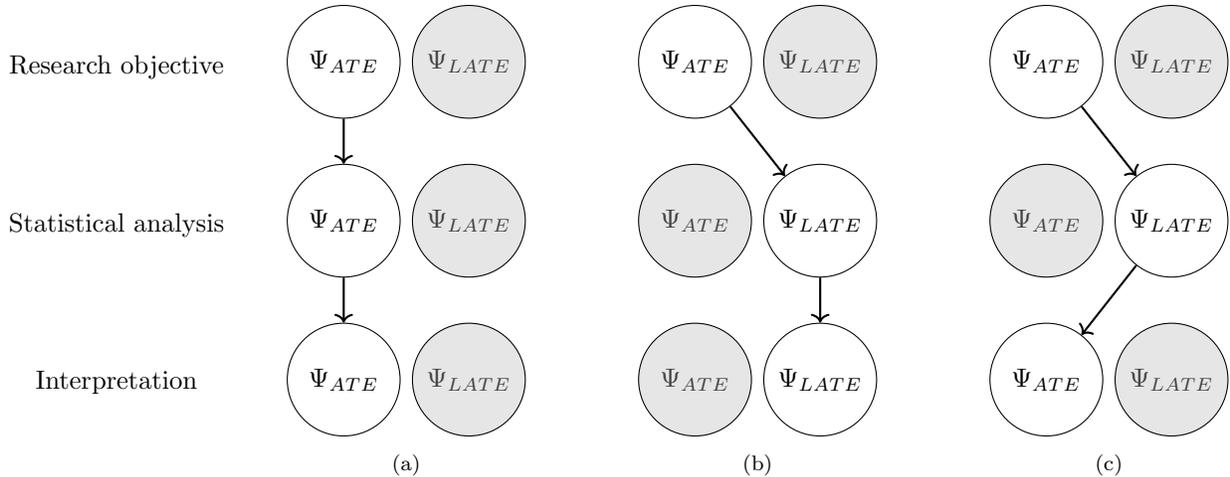

Both strategies have their merits, and we take no normative stance in favor of one over the other. However, we contend that \textit{pragmatic} studies are exposed to a particular source of error. Authors of \textit{pragmatic} studies, seeking consistency with the original research objective, might mistakenly interpret their results in a way that suggests they identified $\ATE$, when in fact they only identified $\LATE$—whose value likely differs, and may even have the opposite sign. We refer to this interpretational error as \textit{identity slippage}. Later in this article, we will formalize this phenomenon to facilitate its empirical measurement. For now, we provide a graphical representation, outlined in Figure \ref{fig:levels}, and two illustrative examples to fix ideas.

As a first example, consider \cite{sequeira_immigrants_2020}, whose abstract begins: ``We study the effects of European immigration to the U.S. during the Age of Mass Migration (1850–1920) on economic prosperity.'' While the authors do not seem to have held a particular interest in $\LATE$, they targeted this parameter in their empirical analysis\footnote{The authors did not make this explicit in the section \textit{Empirical Strategy}, but later wrote that ``A third potential reason for the difference between the two estimates is that 2SLS estimates a local average treatment effect (LATE), which is the average effect amongst compliers, which in our setting are counties whose immigrant population was strongly affected by the presence of a railway'', clearly indicating that they targeted $\LATE$.}, indicating that they followed a \textit{pragmatic} approach. Their abstract continues: ``Exploiting cross-county variation in immigration that arises from the interaction of fluctuations in aggregate immigrant flows and of the gradual expansion of the railway network, we find that counties with more historical immigration have higher income, less poverty, less unemployment, higher rates of urbanization, and greater educational attainment today.'' The claim omits any reference to the local nature of the target estimand and implicitly suggests to readers that $\ATE$ was identified, although, in fact, it was not. This constitutes an instance of \textit{identity slippage}.

% As a second example, consider \cite{acemoglu_weak_2020}, who used rainfall as an instrument to study the effect of Mafia presence in Sicily in 1900 on literacy and other public goods.\footnote{In the same article, the authors used the same instrument to study the effect of the Peasant Fasci movement on Mafia presence.} The authors did not express a specific interest in $\LATE$. However, their stated assumptions do not permit point identification of $\ATE$, and we therefore infer that their target of inference was $\LATE$.\footnote{One might argue that the use of standard regression equations implies an assumption of constant treatment effects. However, in \cite{acemoglu_population_2020}—another article co-authored by one of the same researchers and published in the same journal and year—similar regression equations are used, and the authors explicitly acknowledged that ``OLS and IV estimates capture different local average treatment effects,'' suggesting that the they do not interpret regression equations as an assumption of constant treatment effects.} In the whole article, the authors did not provide a single indication of the local nature of the targeted estimand. When interpreting their results, they wrote: ``Quantitatively, the effects are large. For example, increasing the Mafia in 1900 variable from 1 to 2 [...] will lead to about 10 percentage points decline in literacy in 1921 [...].'' This statement clearly implies that $\ATE$ was targeted and thus constitutes an instance of \textit{identity slippage}.

As a second example, consider \cite{cook_evolution_2023}, who used white casualties in World War II as an instrument to study the effect of racial composition on the share of nondiscriminatory businesses. The authors did not express a specific interest in $\LATE$. However, their stated assumptions do not permit point identification of $\ATE$, and we therefore infer that their target of inference was $\LATE$. In the whole article, the authors did not provide a single indication of the local nature of the targeted estimand. When interpreting their results, they wrote: ``Consistent with this intuition, we find the IV estimate [...] to be substantially larger than the OLS estimates. Here, we see [...] that a 10\% increase in the change in \textit{Share Black} results in a 2.2\% increase in the change in nondiscriminatory shares of formal accommodations, a 0.6\% increase in eating and drinking establishments [...], and a 0.16\% increase in gas station shares [...].'' This statement clearly implies that $\ATE$ was targeted and thus constitutes an instance of \textit{identity slippage}.

\section{Practical relevance of identity slippage}
\label{relevance}

In this section we illustrate practical risks of \textit{identity slippage}. Specifically, we show that generalizations from $\LATE$ to $\ATE$ can be subject to substantial uncertainty that remains undeclared when \textit{identity slippage} occurs. 

In IV analyses, \textit{identity slippage} occurs when a study mistakenly suggests that its results represent estimates of $\ATE$ instead of $\LATE$. Arguably, if $\LATE$ is a good approximation of $\ATE$, meaning that the difference between the true values of the two parameters is small, practical consequences of \textit{identity slippage} are negligible. In contrast, if the difference between the two parameters is large, the consequences may be substantial. To see what factors contribute to this difference, consider again the binary setting assuming monotonicity and notation introduced in Section \ref{identification_iv}. In this setting, the population can be divided into compliers and non-compliers \citep{angrist_identification_1996}. We use the variable $C$ to indicate compliance status. The expected difference between $\LATE$ and $\ATE$, denoted $\Delta$, can be expressed as 
\begin{align}
    \Delta &= E[\LATE - \ATE] \\
    &= P(C=0) \cdot (E[Y(1) - Y(0) \mid C = 1] - E[Y(1) - Y(0) \mid C = 0]),
\end{align}
where the first term represents the probability of non-compliance and the second term the difference in conditional ATEs between compliers and non-compliers. The greater the probability of non-compliance, and the greater the effect heterogeneity between compliers and non-compliers, the larger the gap between $\LATE$ and $\ATE$. For a specific application, the share of non-compliers can be estimated from the data \citep{swanson_challenging_2018}, but the effect heterogeneity cannot, as the conditional ATE among non-compliers remains unidentified. As a result, neither $\ATE$ nor $\Delta$ can be estimated without making additional assumptions. While this means the magnitude of $\Delta$ cannot be determined in practice, we offer an argument suggesting that researchers targeting $\LATE$ implicitly express a belief that this difference is likely meaningful. If a researcher believed there was no treatment effect heterogeneity between compliers and non-compliers, they could simply state this assumption and point identify $\ATE$ directly. For a careful investigator, the fact that they refrain from doing so implies that they suspect this assumption to be incorrect, and thus expect a significant difference between LATE and ATE.

\begin{figure}
    \centering
    \includegraphics[width=0.8\linewidth]{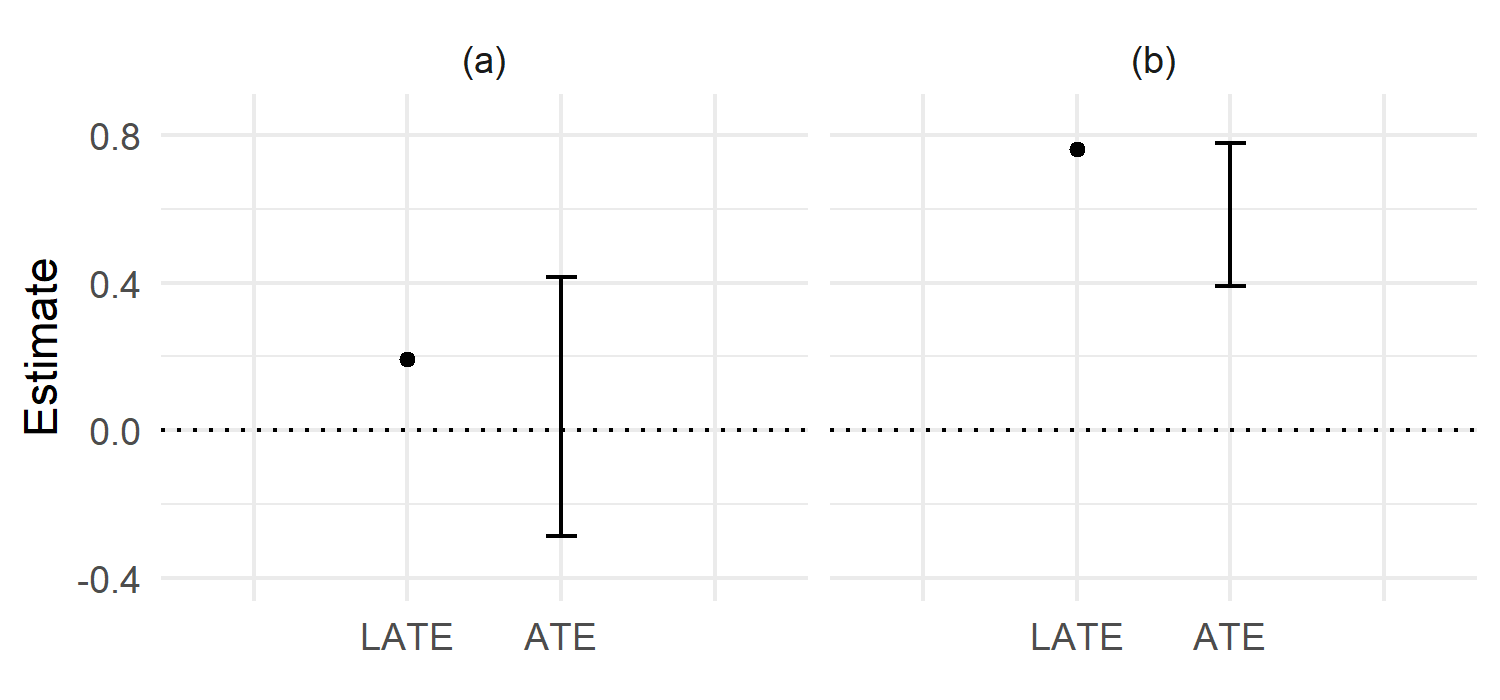}
    \caption{Illustration of examples. Point indicates estimate of $\LATE$ and whiskers indicate sharp bounds of $\ATE$. Panel (a) shows results for \cite{finkelstein_oregon_2012}, and panel (b) shows results for \cite{pearl_causality_2009}.}
    \label{fig:bounds}
\end{figure}

Sharp bounds represent the limits of what can be learned given the data and assumptions. Here, they allow us to quantify the uncertainty involved in generalizing results from $\LATE$ to $\ATE$ for specific applications. The procedure presented in \cite{balke_bounds_1997} identifies sharp bounds of $\ATE$ under the model presented in Section \ref{identification_iv} \citep{swanson_partial_2018}. We applied this procedure to data from \cite{finkelstein_oregon_2012}, who studied the effect of enrollment in Medicaid on healthcare utilization and other outcomes. We estimated the effect of Medicaid on the proportion of participants who visited a primary care physician at least once during the follow-up period, obtaining $\hat{\Psi}_{LATE} = 0.19$ and sharp bounds of $\hat{\Psi}_{ATE} = [-0.29, 0.42]$. We consider a second example from \cite{pearl_causality_2009}, who provides estimates for the effect of cholestyramine on cholesterol levels. The author estimated $\hat{\Psi}_{LATE} = 0.76$ and $\hat{\Psi}_{ATE} = [0.39, 0.78]$. Figure \ref{fig:bounds} illustrates the results of the two examples.

In both examples, $\hat{\Psi}_{LATE}$ was positive. Nevertheless, policy decisions may hinge instead on the sign of $\ATE$, a different parameter. Thus, a stakeholder will rather be interested in what values of $\ATE$ are compatible with the data. In the second example, the sharp bounds $\hat{\Psi}_{ATE}$ were informative about the sign of $\ATE$ (in agreement with $\LATE$). However in the first example, $\ATE$ may be either positive or negative. A study that refrains from inference on the bounds of $\ATE$ would not be able to distinguish between these two very different states of affairs. This demonstrates that an estimate of $\LATE$ alone is insufficient to determine whether the data provide evidence about the sign of $\ATE$. Presenting $\LATE$ as an approximation of $\ATE$ may lead to the mistaken impression that evidence for a positive average effect in the study population was provided, although, in fact, there was not. This shows that presenting results in a way that mistakenly suggests that $\ATE$ was estimated can be misleading. In consequence, \textit{identity slippage} poses a risk to the translation of scientific findings into actionable implications, similar to other common errors or biases that can arise in an analysis.

\section{A systematic review of identity slippage}
\label{review}
Many errors that pose risks to an analysis depend on a discordance between a study's action, and some a priori unknown state of affairs. For example, model misspecification arises from a discordance between an assumed parametric model and the true function relating parameters of the data generating mechanism. Whereas the former is measurable, the latter is not. This challenges the measurement of such errors in scientific practice. In contrast, \textit{pragmatism} and \textit{identity slippage} depend on a discordance between two different actions of a study. As such, these phenomena are amenable to statistical description, which we do here. 

To quantify their prevalence in the applied IV literature, we followed preregistered study protocol, available  on \href{https://osf.io/fxq5v}{OSF} under the registration id `fxq5v'. Specifically, we conducted a systematic review of articles published in leading journals of economics, political science, epidemiology and medicine. In what follows, we present a formalization of these phenomena, which is tailored to IV analysis and enables transparent and reproducible measurement in empirical studies. For a formalization that applies to a generalized context, we refer to \cite{sarvet_interpretational_2023}. We then discuss details on sample construction and present the results of our study.

\subsection{Definitions and measurement}
\label{review_definitions}

As illustrated in Figure \ref{fig:levels}, \textit{pragmatism} and \textit{identity slippage} are defined in terms of estimands related to a study's research objective, statistical analysis, and result interpretations. In our formalization, we represent each of these levels with a variable that is amenable to measurement in empirical studies. As studies rarely report the values of these variables directly, we defined a strategy to measure them as specific and transparent functions of other elements routinely stated in scientific articles. We then deployed these functions within the population of included studies. To mitigate potential reviewer disagreement, we specify these functions as an exhaustive set of rules and defining examples designed to unambiguously classify article elements into estimands of different types. As a caveat, we acknowledge that others may reasonably specify these functions differently.\footnote{In the Appendix, we provide sensitivity analyses where we consider alternative specifications of these functions.}

\subsubsection{Estimand of interest $\estimandOfInterest$}
Most empirical studies begin with an articulation of a research objective. Here we suppose that there exists a function that maps a research objective to an estimand of interest. We specify this function in order to measure a study's estimand of interest from their research objectives, which are often stated in plain language. To do so, we let $\mathfrak{O}$ be the set of all possible causal research objectives and let $\Gamma$ denote set of all possible estimands. To simplify matters, we consider the subset of estimands that are commonly targeted by IV strategies, denoted $\Gamma_{IV} = \{\LATE, \ATE\} \subset \Gamma$. Suppose that every study that uses IV to estimate causal effects has a research objective $O \in \mathfrak{O}$ and the function $\Omega: \mathfrak{O} \mapsto \Gamma_{IV}$ translates causal research objectives into causal estimands according to some (potentially highly-contingent) socially-constructed scientific norms. We define the \textit{estimand of interest} $\estimandOfInterest := \Omega(O)$ to be the causal estimand whose estimation would address the research objective of the study. As an example, consider a study whose authors are interested in the average effect of some treatment in the study population. One such mapping $\Omega$ would assert that this research objective would be addressed through the estimation of $\ATE$ and $\estimandOfInterest = \ATE$.\footnote{In this analysis, we take for granted a particular mapping $\Omega$; therefore, in what follows, we restrict our considerations to the specification of that mapping. However, we acknowledge the specification of $\Omega$ could be contested, or vary across communities of scientists, policy makers, and other stakeholders or consumers of research. Furthermore, one might disagree that such an $\Omega$ even exists at all.} 

Measurement of $\estimandOfInterest$ in empirical studies requires knowledge of $O$ and a generally applicable specification of $\Omega$. We measure $O$ by extracting the articulation of a study's research objective, typically from its abstract. We specify $\estimandOfInterest := \Omega(O) = \LATE$ if the extracted excerpt indicates a specific interest in $\LATE$. We consider two types of statements that indicate interest in $\LATE$. Either, authors state an interest in the sub-population for which $\LATE$ is valid, or in the effect of treatment changes induced by the instrument. Otherwise, we specify $\estimandOfInterest = \ATE$.

As examples, consider the following research objectives extracted from empirical studies:
\begin{itemize}
    \item ``This article studies the effect of corporate and personal taxes on innovation in the United States over the twentieth century.'' \citep{akcigit_taxation_2022}
    \item ``This article tests for bias in consumer lending using administrative data from a high-cost lender in the U.K. [...] Though IV estimators are often criticized for the local nature of the estimates, we exploit the fact that the outcome test relies on the difference between exactly these kinds of local treatment effects to test for bias in consumer lending.'' \citep{dobbie_measuring_2021}
\end{itemize}

In the first example, authors did not indicate that they held a specific interest in $\LATE$, and our specification implies $\estimandOfInterest = \ATE$. In the latter example, authors motivated their interest in the local nature of $\LATE$ and our specification implies $\estimandOfInterest = \LATE$.

\subsubsection{Targeted estimand $\targetedEstimand$}
Empirical applications of IV involve a statistical analysis, comprised by a specified model and a set of analytic procedures. Here we suppose that there exists a function that maps a study's statistical analysis to its targeted estimand. We specify this function in order to measure a study's targeted estimand from their statistical analysis, which are often described in methods sections and other supporting materials.

Let $\mathfrak{S}$ denote the set of all possible statistical analyses. Suppose that every study has a statistical analysis $S \in \mathfrak{S}$ and a function $\Xi: \mathfrak{S} \mapsto \Gamma_{IV}$ exists.\footnote{We expect the existence of such a function $\Xi$ is not a controversial supposition, and in many cases would simply be evaluated by examining the properties of the study's analytic procedures under the study's assumed model, e.g., the causal parameter that an estimator is consistent for under a conventional asymptotic regime.} We define the \textit{targeted estimand} as $\targetedEstimand := \Xi(S)$. As an example, consider an IV analysis that estimates a causal effect under the assumptions \eqref{ass:1} - \eqref{ass:5} presented in Section \ref{identification_iv}. Under these assumptions, this study's estimator is consistent for $\LATE$ but not $\ATE$, and our specification implies $\targetedEstimand = \LATE$. Alternatively, consider an IV analysis with estimators consistent for sharp upper and lower bounds $\ATE$. In this case $\targetedEstimand = \ATE$.

We measure $S$ by extracting the description of the statistical analysis from a given study. Specifically, we classify $\targetedEstimand := \Xi(S)= \ATE$ if the authors either made explicit homogeneity assumptions that allow for point identification of $\ATE$, or targeted bounds of $\ATE$. Otherwise, we classify $\targetedEstimand = \LATE$. For studies where $\targetedEstimand = \LATE$, we determine whether the authors explicitly specified that their target estimand is $\LATE$.\footnote{Note that we did not consider linear regression equations as homogeneity assumptions, unless there was a clear indication that the regression equation represents the presumed data generating process. If one is willing to interpret a regression equation as the presumed data generating process, a linear regression equation (without interactions) implies constant (and therefore homogeneous) treatment effects. Here, we considered a regression equation merely as a representation of conditional outcome means in the factual data, which does not automatically imply treatment effect homogeneity. However, we performed a sensitivity analysis, in which we considered linear regression equations to serve as explicit homogeneity assumptions.}

As examples, consider the following descriptions:
\begin{itemize}
    \item ``We derive conditions for this instrument to identify $\lambda$ in a simplified setting where [...] there is no unobserved treatment effect heterogeneity [...].'' \citep{abaluck_mortality_2021}
    \item ``The parameter of interest is $\beta_{2}$, which identifies the local average treatment effects (LATEs) of military service among individuals who were near the applicable AFQT cutoff and were induced to serve or not serve in the military based on their position relative to their cutoffs.'' \citep{greenberg_army_2022}
    \item ``To demonstrate the causal effect of government spending on patriotic actions, we use arguably exogenous variation in New Deal support.'' \citep{caprettini_new_2023}
\end{itemize}

In the first example, authors made an explicit homogeneity assumption that allows for point identification of $\ATE$, and our specification of $\Xi$ implies $\targetedEstimand = \ATE$. In the second example, authors explicitly specified $\LATE$ as their target parameter, and our specification of $\Xi$ implies $\targetedEstimand = \LATE$. In the third example, authors did not specify their target parameter explicitly. However, they also did not state assumptions that allow for point identification of $\ATE$, and our specification of $\Xi$ implies $\targetedEstimand = \LATE$.

\subsubsection{Claimed estimand $\claimedEstimand$}
Articles describing empirical applications nearly always include plain language interpretations of results or claims. Ideally, interpretations would be appropriate given the study's targeted estimand, but sometimes a claim or interpretation would only be appropriate if a different estimand had been targeted. We refer to this as the study's \textit{claimed estimand}.  Here we suppose that there exists a function that maps a set of a study's plain language claims or interpretations to its claimed estimands. We specify this function in order to measure these claimed estimands from a study's interpretations, which are often described in results, discussions and abstract sections. 

Let $\mathfrak{I}$ be the set of all possible claims and scientific implications, expressed in natural language. Let $P(\mathfrak{I})$ denote the power set of $\mathfrak{I}$. Suppose that every study makes a non-empty set of claims $C$ and a function $\Phi: P(\mathfrak{I}) \mapsto \Gamma_{IV}$ exists. We define $\claimedEstimand = \Phi(C_j)$ to be the \textit{claimed estimand} corresponding to $C_j$, the subset of claims in $C_j\subseteq C$. As an example, consider a study that claims that their results show that a treatment has on average a positive effect in the study population. This claim is only appropriate if $\ATE$ is targeted and $\claimedEstimand = \ATE$. Alternatively, consider a study that claims that their results show that a treatment has on average a positive effect in the sub-population of compliers. This claim is only appropriate if $\LATE$ is targeted and thus $\claimedEstimand = \LATE$.

For each article section that potentially contains result interpretations, we determine $\claimedEstimand$ individually. We consider the sections \textit{Abstract} ($\claimedEstimandAbstract$), \textit{Results} ($\claimedEstimandResults$) and \textit{Discussion} ($\claimedEstimandDiscussion$).\footnote{We standardize sections to `Abstract`, `Results`, and `Discussion`. If a study has a different structure, we consider the reporting and interpretation of results as \textit{Results} and the discussion of results and conclusions as `Discussion`.} For each of these sections, we extract the set of claims $C_s$ related to the IV analysis. If $C_s$ is empty, i.e. no claims related to the IV analysis are made, $\claimedEstimand^s$ remains unclassified. If $C_s$ is non-empty, we decide for each claim whether it is appropriate for the targeted estimand or for some other estimand. If $\LATE$ is targeted and all claims in $C_s$ are appropriate for $\LATE$, we classify $\claimedEstimand^s := \Phi(C_s)= \LATE$. Otherwise, we classify $\claimedEstimand^s := \Phi(C_s)= \ATE$.

To apply $\Phi$ in the context of this study, we give a precise definition of what is and is not an appropriate interpretation for $\LATE$. To illustrate, consider the canonical IV setting with an instrument $Z$, a treatment $A$ and an outcome $Y$. Suppose authors use an IV to target $\LATE$ and estimate an effect of magnitude $M > 0$. Suppose that in a supplementary analysis, the authors also target $\ATE$ through covariate adjustment and estimate an effect of magnitude $L$. We distinguish between six types of claims. These claim types were developed during a pilot analysis and specified prior to the construction of the final sample. Table \ref{table:claim-types} lists the claim types and provides examples.
\begin{table}[h!]
\begin{center}
\def\arraystretch{1.05}
\begin{tabular}
{|p{0.075\textwidth}|p{0.3\textwidth}|p{0.5\textwidth}|}
    \hline
    Type &  Structure & Example \\
    \hline
    a) & ``A has an effect of magnitude M on Y for compliers.'' & ``Compliers add an average of 1.6 schools to their application list, and reduce their ex post nonplacement risk by 15.5 percentage points, equal to 58\% of the below-threshold mean..'' \citep{arteaga_smart_2022}  \\
    \hline
    b) & ``Changes in A induced by Z have an effect of magnitude M on Y.'' &  ``Column 1 of panel A reports a 2SLS point estimate of 0.989, indicating that allowances induced by judge leniency increase DI receipt almost one-for-one in the first year following appeal.'' \citep{autor_disability_2019}  \\
    \hline
    c) & ``A has a positive effect on Y.'' &  ``We [...] find support for a causal interpretation of the relationship between legal change, human capital, and growth.'' \citep{dittmar_public_2020}  \\
    \hline
    d) & ``A has an effect of magnitude M on Y.'' &  ``Quantitatively, the effects are large. For example, increasing the Mafia in 1900 variable from 1 to 2 [...] will lead to about 10 percentage points decline in literacy in 1921 [...].'' \citep{acemoglu_weak_2020} \\
    \hline
    e) & ``In our study population A has on average a positive effect on Y.'' &  ``[...] we find that counties with more historical immigration have higher income, less poverty, less unemployment, higher rates of urbanization, and greater educational attainment today.'' \citep{sequeira_immigrants_2020} \\
    \hline
    f) & ``The difference between M and L suggest that L is (not) subject to confounding.'' &  ``We also find little difference between IV and OLS estimates [...], indicating that the effects of potential endogeneity of inflation expectations with respect to firms’ price setting decisions are limited.'' \citep{coibion_inflation_2020} \\
    \hline
\end{tabular}
\end{center}
\caption{Claim types with examples}
\label{table:claim-types}
\end{table}

Claims of type (a) clarify that the estimates are valid only for a sub-population. Claims of type (b) clarify that the estimates are only valid for a certain type of change in A, namely the change induced by Z. Claims of type (c) state that there is a positive effect of A on Y. This statement is technically correct if at least one unit experiences a positive effect, which is implied by $M > 0$. We specify claims (a) - (c) to be appropriate given that $\LATE$ was targeted. Claims of type (d) interpret the magnitude of the average effect but do not clarify that the effect is only valid for a sub-population. Claims of type (e) make a reference to the average effect in the study population. Claims of type (f) compare the IV estimate $M$ with the covariate adjustment estimate $L$ that targets $\ATE$. We assert that claims (d) - (f) implicitly suggest that $\ATE$ was targeted by the IV analysis. Thus, we specify these claims to be inappropriate given that $\LATE$ was targeted.\footnote{Note that we consider claims of type (a) or (b) as clear references to the local nature of $\LATE$. If a claim of type (a) or (b) occurs in a section, we consider all other claims in the section to be of type (a) or (b).} As an example, suppose that $C$ consists of claims of type (a), (b) or (c). Then, according to this specification, $\claimedEstimand = \Phi(C) = \LATE$. Similarly, if $C$ contains claims of types (d), (e) or (f), then $\claimedEstimand = \Phi(C) = \ATE$.

\subsubsection{A pragmatic approach and identity slippage}
Based on these variables, we can now directly measure whether studies are \textit{strict} or \textit{pragmatic}, and whether they contain instances of \textit{identity slippage}. We specify that concordance between $\estimandOfInterest$ and $\targetedEstimand$ indicates a \textit{strict} study and discordance indicates a \textit{pragmatic} study. For all studies with $\targetedEstimand = \LATE$, we evaluate each section separately to determine whether \textit{identity slippage} occurs therein. We specify that discordance between $\claimedEstimand$ and $\targetedEstimand$ indicates \textit{identity slippage}.

\subsection{Data}
\label{review_data}
We identified the population of IV applications published between 2019 and 2023\footnote{We consider publication dates, not issue dates.} in major journals of the economic, political, epidemiological, and medical sciences. The included journals are listed in Table \ref{table:included-journals}. We used Dimensions, a fee-based research database, to construct our sample by targeting the selected journals and running a full-text search for `instrumental variable`.

\begin{table}[h]
\begin{center}
\begin{tabular}{|l|l|}
\hline
    Economics & American Economic Review \\
    & Econometrica \\
    & Journal of Political Economy \\
    & Quarterly Journal of Economics \\ 
    & Review of Economic Studies \\
    \hline 
    Political science & American Political Science Review \\
    & American Journal of Political Science \\
    & Journal of European Public Policy \\
    & The Journal of Politics \\
    & British Journal of Political Science \\
    \hline
    Epidemiology & Epidemiology \\
    & American Journal of Epidemiology \\ 
    & International Journal of Epidemiology \\
    & European Journal of Epidemiology \\
    & Annals of Epidemiology \\
    \hline
    Medicine &  The Lancet \\
     & New England Journal of Medicine \\
     & JAMA \\
     & The BMJ \\
     & Nature Medicine \\
    \hline
\end{tabular}
\end{center}
\caption{Journals included in sample}
\label{table:included-journals}
\end{table}

We manually screened all returned studies using the following inclusion criteria:
\begin{itemize}
    \item IV applied in original empirical investigation
    \item IV results presented in main article
    \item Research objective is identification of a causal effect
\end{itemize}

The first criterion excludes other article types, for example methodological articles or systematic reviews, and studies that mention but do not apply IV. The second criterion excludes studies that present their IV results exclusively in an appendix. The third criterion excludes applications where estimating a narrowly defined causal effect is not the primary research objective - for instance, when researchers use instrumental variables (IV) to estimate parameters of a structural model.\footnote{We focus on applications that \cite{angrist_instrumental_2001} describe in the section `Instrumental Variables and Omitted Variables'. As an example of a study that was excluded, see \cite{duranton_urban_2023}. In this study, authors develop a structural model and use IV to estimate a parameter of this model. We assert that in such settings, authors are unlikely to interpret their results as `treatment effects'.}

For each included study, the first author of this article classified the set of variables $\{\estimandOfInterest, \targetedEstimand, \claimedEstimandAbstract, \claimedEstimandResults, \claimedEstimandDiscussion\}$ following the specification outlined in the previous section. Based on the measured variables, we derived whether a study was \textit{strict} or \textit{pragmatic}. For studies with $\targetedEstimand = \LATE$, we evaluated, separately for each section, whether \textit{identity slippage} occurred.

\subsection{Results}
\label{review_results}

We constructed our sample as specified in the previous section. The entire data set will be made publicly available as indicated in the Appendix. Figure \ref{fig:article-flow} illustrates the sample selection process. Our query returned a total of 854 studies, 309 of which met the inclusion criteria. Of the 309 studies in our sample, 93 (30\%) were published in economic, 61 (20\%) in political science, 114 (37\%) in epidemiological, and 41 (14\%) in medical journals. Across all disciplines, instrumental variable (IV) methods were used in both natural experiments (n~=~276, 89\%) and randomized controlled trials (RCTs) (n~=~33, 11\%). In what follows, we examine the prevalence of \textit{pragmatism} and \textit{identity slippage} and compare patterns observed in economics with those in other disciplines. Sensitivity analyses and a more comprehensive cross-discipline comparison are provided in the Appendix.

\tikzstyle{box} = [rectangle, rounded corners, minimum width=6.5cm, minimum height=2cm, text centered, text width=6.5cm, draw=black, fill=white]
\tikzstyle{arrow} = [thick,->,>=stealth]
\begin{figure}[t!]
    \begin{center}
    \begin{tikzpicture}[node distance=3.3cm]
    
    \node (query) [box] {Studies returned from query (n = 854)};
    \node (screened) [box, below of=query] {Studies screened (n = 854)};
    \node (excluded) [box, right of=screened, node distance=8.3cm, inner sep=5pt] {Studies excluded (n = 545):
    {\footnotesize
    \begin{itemize}
        \item No IV application (n = 465)
        \item IV not in main article (n = 40)
        \item Other research objective (n = 40)
    \end{itemize}}};
    \node(analysis) [box, below of=screened, node distance = 3.6cm, inner sep=5pt]
    {
    Studies included in analysis\\(n = 309):
    {\footnotesize
    \begin{itemize}
        \item Economics (n = 93)
        \item Political science (n = 61)
        \item Epidemiology (n = 114)
        \item Medicine (n = 41)
    \end{itemize}}};
    
    \draw [arrow] (query) -- (screened);
    \draw [arrow] (screened) -- (excluded);
    \draw [arrow] (screened) -- (analysis);
    
    \end{tikzpicture}
    \end{center}
    \caption{PRISMA flow diagram illustrating sample selection process}
    
    \label{fig:article-flow}
\end{figure}
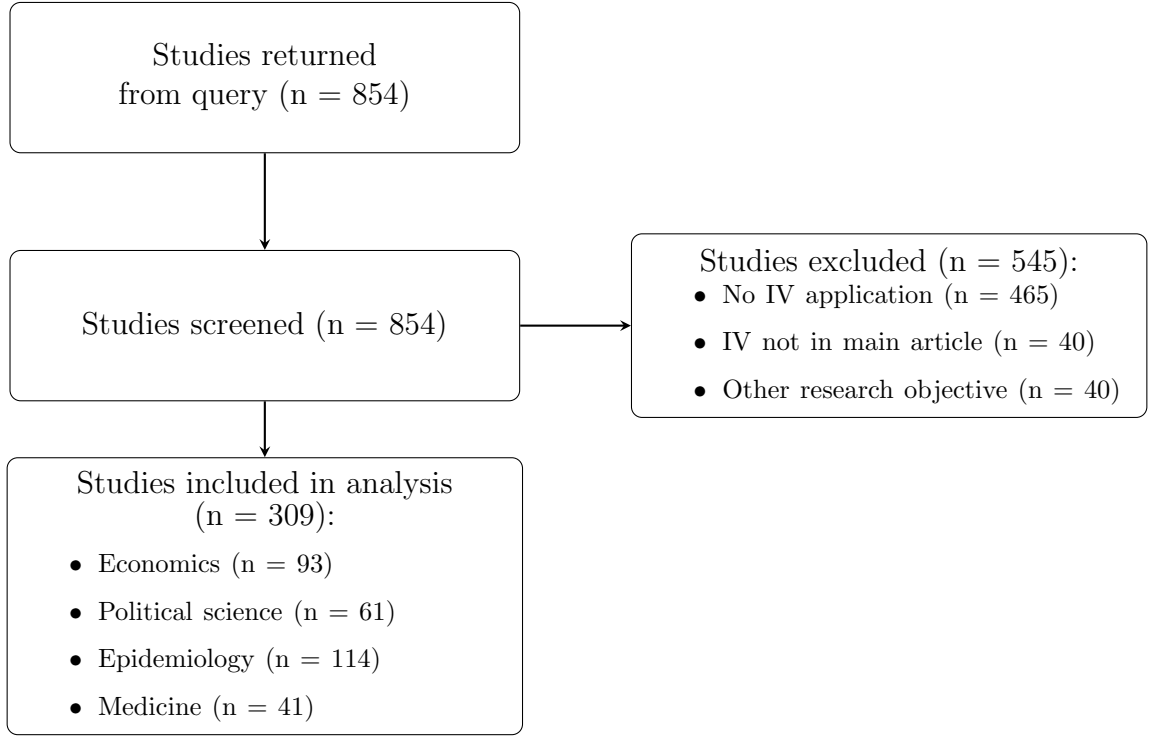

\begin{figure}[h]
    \includegraphics[width=\textwidth]{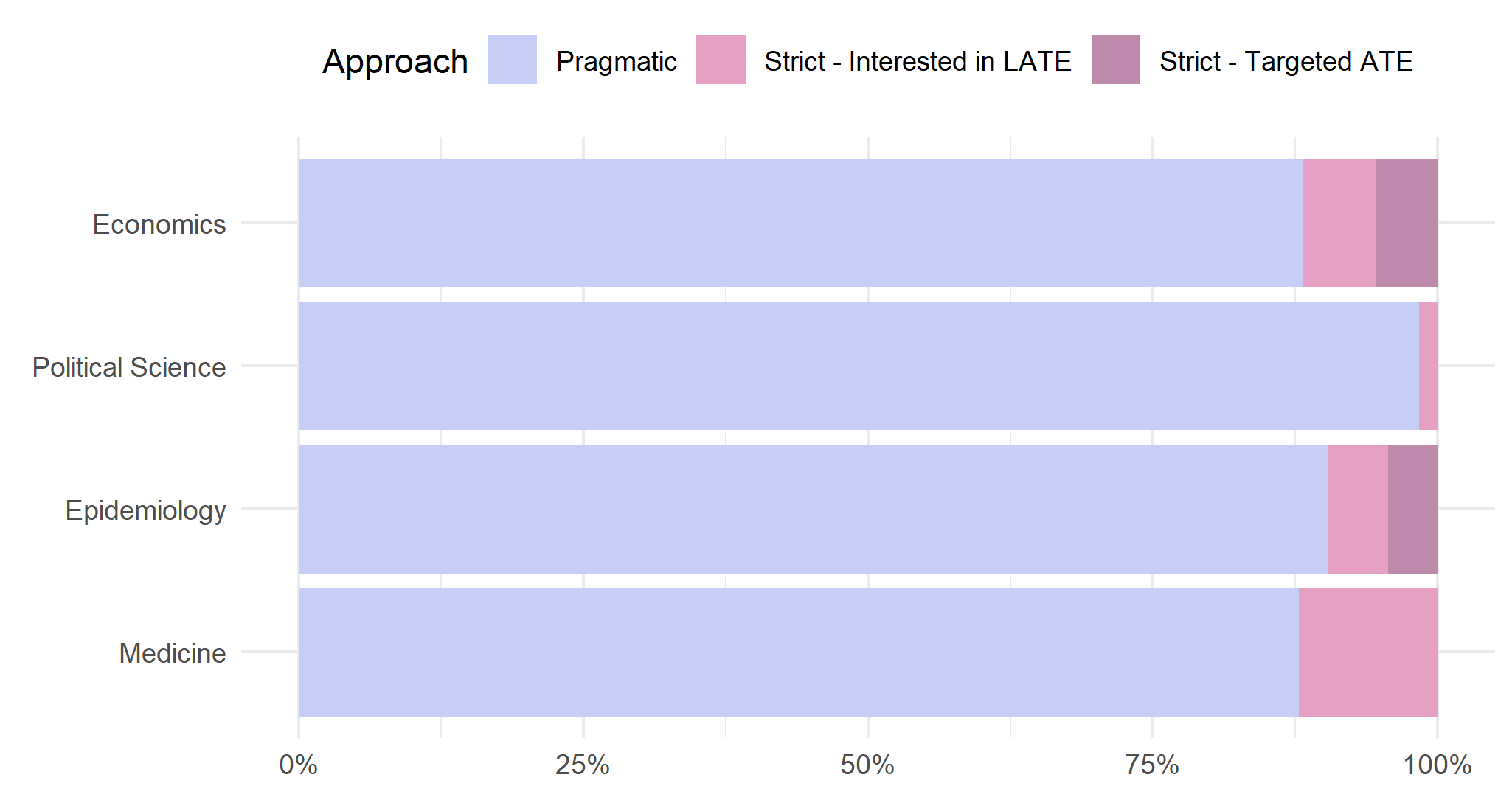}
\caption{Prevalence of the pragmatic approach by discipline.}
\label{fig:results_framework}
\end{figure}

The majority of studies were classified to have taken a \textit{pragmatic} approach (n~=~281, 91\%). Both studies that expressed an explicit interest in $\LATE$ (n~=18, 6\%) or targeted $\ATE$ (n~=~10, 3\%) were rare. Notably, none of the studies in our sample targeted bounds of $\ATE$. Figure \ref{fig:results_framework} illustrates the distribution of \textit{pragmatic} and \textit{strict} studies across disciplines. Within economics, 82 studies (88\%) were classified to have taken a  \textit{pragmatic} approach. As in the overall sample, both studies that expressed an explicit interest in $\LATE$ (n~=~6, 6\%) or targeted $\ATE$ (n~=~5, 5\%) were rare. The observed proportion of \textit{strict} studies was consistently low across disciplines, with political science exhibiting the lowest rate (n~=~1, 2\%). 

All studies that targeted $\LATE$ were evaluated for \textit{identity slippage} (n~=~299, 97\%). Of these, 183 (61\%) exhibited instances of \textit{identity slippage} in at least one article section, and 145 (48\%) omitted any explicit reference to the local nature of the estimand, leaving it to readers to infer. Prevalence of \textit{identity slippage} was lower in studies that explicitly referenced the local nature of $\LATE$ in the description of the statistical methodology (n~=~23, 33\%), and in studies classified as \textit{strict} (n~=~2, 11\%). Natural experiments (n~=~175, 65\%) exhibited a higher prevalence than RCTs (n~=~8, 27\%). A substantial proportion of studies did not contain any IV related claims in \textit{Discussion} (n~=~42, 14\%) or \textit{Abstract} (n~=~52, 17\%), and for these sections, \textit{identity slippage} was not evaluated. Of all evaluated sections, 319 (40\%) were affected by \textit{identity slippage}. Most frequently, \textit{identity slippage} occurred in \textit{Results} (n~=~138, 46\%), followed by \textit{Abstract} (n~=~103, 34\%) and \textit{Discussion} (n~=~78, 26\%).\footnote{Proportions were calculated with respect to number of studies for which \textit{identity slippage} was evaluated.}

\begin{figure}[t]
    \includegraphics[width=\textwidth]{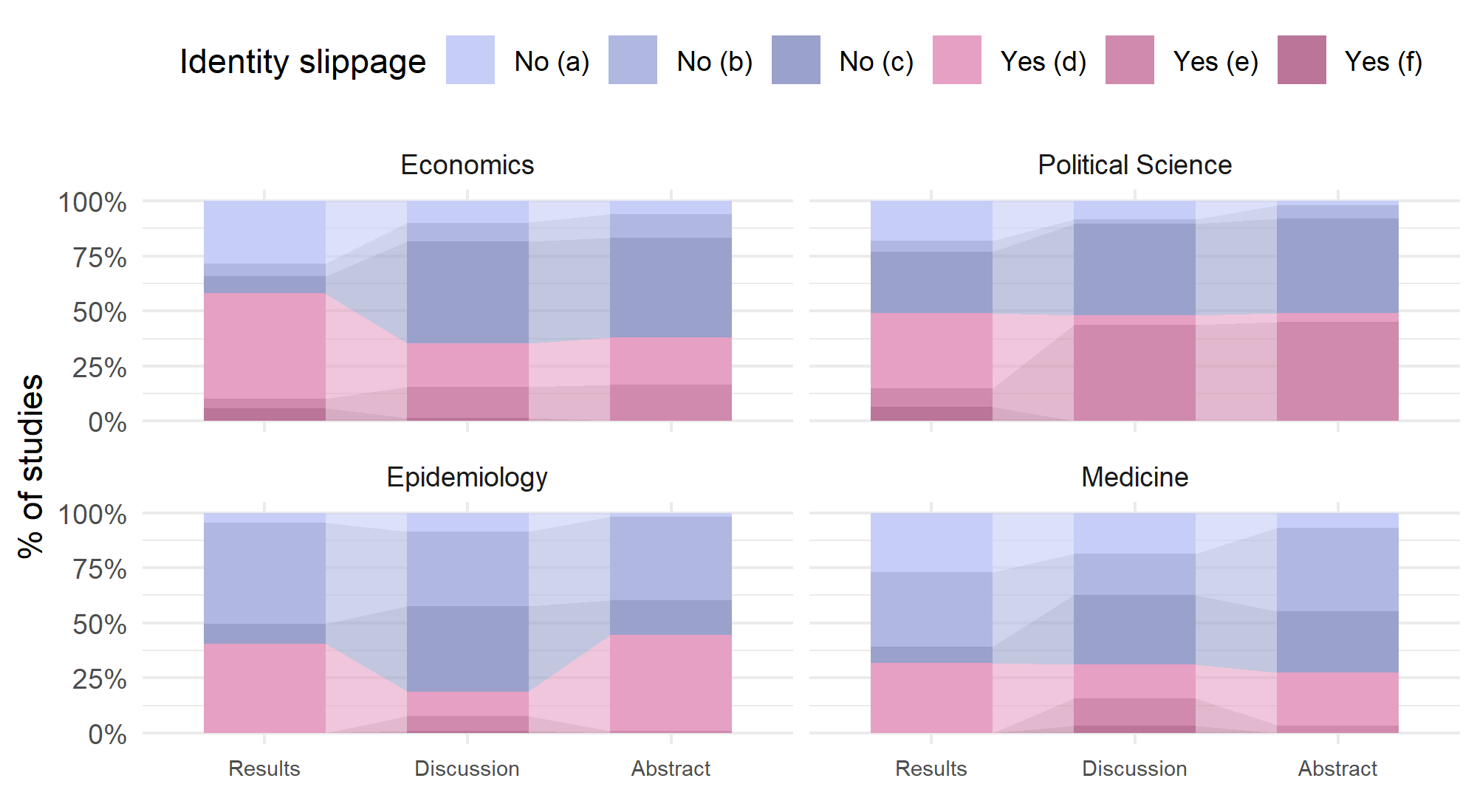}

\caption{Prevalence of identity slippage across article section by discipline. Letter in parentheses denotes claim type.}
\label{fig:results_slippage}
\end{figure}

Figure \ref{fig:results_slippage} illustrates the prevalence of \textit{identity slippage} and distribution of claim types across sections and disciplines. We found \textit{identity slippage} to be widespread across disciplines. Within economics, \textit{identity slippage} occurred at least once in 60 studies (68\%). Quantitatively, prevalence was comparable in political science science (n~=~42, 69\%), and somewhat lower in epidemiology (n~=~61, 56\%) and medicine (n~=~22, 46\%). However, the nature of interpretations differed substantially across fields. For example, type (a) claims—arguably the clearest reference to the local nature of $\LATE$—were far more common in economics than in epidemiology, where type (b) claims predominated. For this reason, we refrain from comparing the overall severity of \textit{identity slippage} across disciplines solely on the basis of the presented numbers. Rather, we conjecture that it affects a substantial share of studies in each field. A more detailed cross-discipline analysis is provided in the Appendix.

\section{Mechanisms}
\label{susceptibility}
In the previous section, we have shown that \textit{identity slippage} is widespread in IV applications that target $\LATE$. In this section, we discuss potential mechanisms that can lead to \textit{identity slippage}.

The estimation of $\LATE$ was most often accompanied by a \textit{pragmatic} decision to abandon the original parameter of interest. As we have argued in Section \ref{identity_slippage}, this creates fertile ground for \textit{identity slippage}, and our empirical results are consistent with this hypothesis. In \textit{pragmatic} studies, \textit{identity slippage} can arise through two pathways. First, investigators who aim to preserve a coherent narrative may consciously decide to present their results in a way that is consistent with their original research objective, and not with their statistical analysis. Second, investigators might simply be unaware of practical differences between $\LATE$ and $\ATE$ and thus treat them as two exchangeable entities. A second factor that may strengthen the link between  \textit{pragmatism} and \textit{identity slippage} is related to the interpretation of $\LATE$: In settings where the interpretation is particularly challenging, the likelihood of \textit{identity slippage} may increase. On one hand, investigators who prioritize a coherent narrative face an even stronger incentive to obscure the complexity behind the targeted estimand. On the other, we suspect that as interpretation grows more complex, fewer investigators fully understand what implications their target estimand entails.

\begin{figure}[t]
\centering
\begin{tikzpicture}[
  node distance=1cm and 2cm,
  every node/.style={minimum height=1cm, minimum width=3cm, align=center},
  ->, >=Stealth
]

% Nodes
\node (A) at (0,0) {Pragmatism};
\node (B) [below=of A, xshift=1cm] {Interpretability of $\targetedEstimand$};
\node (C) [right=of A] {Identity slippage\\by authors};
\node (D) [right=of C] {Identity slippage\\by readers};

% Edges
\draw (A) to (C);
\draw (A) to[bend right=20] (D);
\draw (B) to (C);
\draw (B) to[bend right=20] (D);
\draw (C) to (D);

\end{tikzpicture}
\caption{Causal diagram illustrating relationships among pragmatism, interpretability, and identity slippage}
\label{fig:driving_factors}
\end{figure}
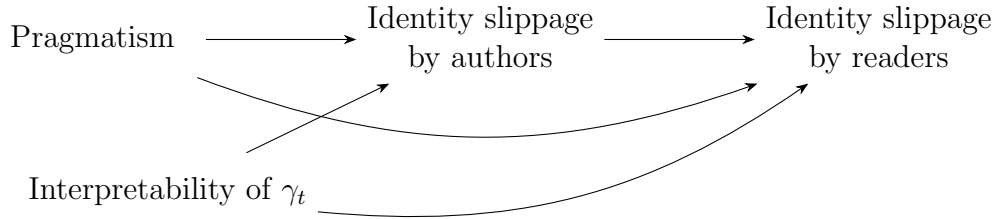

While our empirical analysis is limited to \textit{identity slippage} by authors, we now also briefly discuss its relation to \textit{identity slippage} by readers. Conceptually, extending our definition to readers’ interpretation of results requires substituting $\claimedEstimand$ with a different variable that reflects the estimand a reader perceives the results to represent. \textit{Identity slippage} by authors and readers can, but need not, coincide. A well-informed reader might interpret results correctly, regardless of how authors have presented them. Conversely, a reader might misinterpret results, even if authors did not make any inappropriate claims. We conjecture that \textit{identity slippage} by authors leads to \textit{identity slippage} by readers, at least in tendency if not always in fact. Moreover, we consider it to be plausible that \textit{pragmatism} and the interpretability of $\targetedEstimand$ have effects on \textit{identity slippage} by readers that are not mediated by \textit{identity slippage} by authors. Arguably, in \textit{pragmatic} settings, readers may be inclined to interpret results in a manner consistent with $\estimandOfInterest$, regardless of whether authors have committed \textit{identity slippage}. Similarly, when $\targetedEstimand$ is difficult to interpret, readers may misinterpret the results, regardless of how authors have presented them. Therefore, \textit{pragmatism} and subtleties in the interpretation of the targeted estimand can affect readers' interpretations, even if the authors' interpretations are accurate. Figure \ref{fig:driving_factors} illustrates the discussed relationships.

\section{Discussion}
\label{discussion}
In certain contexts, IV may represent the only feasible approach for identifying causal effects under credible assumptions, thereby expanding the space of relationships for which causal inference can be conducted. However, the use of IV is often accompanied by a pragmatic decision to abandon the identification of original parameter of interest and target $\LATE$ instead. In this article, we suppose that this shift poses an increased risk for interpretational errors. Our systematic review provides some evidence for this claim. We found that while the vast majority of IV applications target $\LATE$, specific interest in this parameter is rare, and interpretational errors were widespread across disciplines. Our findings suggest that the validity of conclusions drawn from results of IV applications could often be compromised by interpretational errors.

Evaluating the trade-off between the benefits and risks of pragmatic IV estimation is context-specific and, to some degree, subjective. If $\LATE$ is indeed the parameter that is of primary interest to investigators, then it constitutes a meaningful target of inference. While these cases undoubtedly exist, we found them to be rare. If investigators are interested in a different parameter, such as $\ATE$, the relevance of $\LATE$ hinges on the proximity to the parameter of interest. This proximity may be subject to considerable uncertainty in practice, and is important to investigate. Some strategies to quantify such proximity rely on partial identification approaches, profiling of compliers, or on an argumentative basis (\cite{swanson_partial_2018}, \cite{swanson_commentary_2013}, \cite{marbach_profiling_2020}). Yet, these approaches were rarely, if ever, used in the articles studied in our review. To minimize risks of misinterpretation, investigators must be aware of the discrepancy between research objective and identification strategy and communicate it transparently. This transparency was often lacking in the articles we studied. Settings in which the interpretation of $\LATE$ is particularly challenging, for example if the instrument is non-causal, put results at further risk of being misinterpreted. 

\cite{imbens_better_2010} defends the use of IV in an article titled ``Better LATE Than Nothing''. As empirical scientists, we want to ground our inferences in data, but must often acknowledge that data alone often cannot provide a satisfying answer to practical questions. In this sense, we agree with his statement, but believe it should not be accepted unconditionally. We argue that, ideally, an IV should be used to provide inference on parameters researchers hold a genuine interest in. When this is not possible, transparency is essential to reduce the risk of misinterpretation.

\newpage
\bibliographystyle{unsrtnat}
\bibliography{references}

\newpage
\section{Appendix}
\label{appendix}
\subsection{Data availability}
Upon publication, the entire data set used for the analysis will be made publicly available in a form compatible with the data sharing policies of the respective journal.

\subsection{Additional results}
Here, we provide a more detailed comparison of the different disciplines and application types. Moreover, we provide sensitivity analyses where we consider alternative specifications of potentially controversial classification criteria.

\subsubsection{Comparison of disciplines}
Prevalence of \textit{identity slippage} was highest in political science (n~=~42, 69\%), followed by economics (n~=~60, 68\%), epidemiology (n~=~61, 56\%) and medicine (n~=~22, 46\%). Table \ref{table:comparison} illustrates the prevalence of \textit{identity slippage} stratified by discipline, approach (pragmatic vs. strict), and whether the estimand was explicitly specified in the description of the statistical methodology.
\begin{table}[h]
    \begin{tabular}{lllll}
    \toprule
                     &           Economics &        Political Science &       Medicine &             Epidemiology \\
    \midrule
          Framework - Strict &      0/11 (0\%) &       0/1 (0\%) &      1/5 (20\%) &    1/11 (9.09\%) \\
       Framework - Pragmatic & 60/82 (73.17\%) &    42/60 (70\%) &    18/36 (50\%) & 61/103 (59.22\%) \\
    Estimand specified - Yes &     5/20 (25\%) &  8/13 (61.54\%) &  2/15 (13.33\%) &   8/24 (33.33\%) \\
     Estimand specified - No & 55/73 (75.34\%) & 34/48 (70.83\%) & 17/26 (65.38\%) &     54/90 (60\%) \\
    \bottomrule
    \end{tabular}
\caption{Prevalence of \textit{identity slippage} by discipline, framework and whether estimand was explicitly specified}
\label{table:comparison}
\end{table}

\subsubsection{Comparison of application types}
We distinguish between three different types of IV applications: RCTs, Mendelian Randomization (MR)\footnote{I.e. genetic variants were used as instruments}, and all other applications. Since MR application occurred almost exclusively in epidemiology and medicine, we limited the discussion in the main article to RCTs and natural experiments in general. Here, we provide a more detailed comparison.

Application of MR accounted for 110 studies (35\%), applications in randomized controlled trials (RCTs)\footnote{I.e. the investigators controlled the assignment of the instrument.} accounted for 33 studies (11\%), and other applications accounted for 166 studies (54\%). Application type was a strong predictor of discipline. MR applications were responsible for a large proportion of studies in epidemiology (n~=~89, 78\%) and medicine (n~=~20, 49\%). RCTs were present in all disciplines, contributing a considerable share of studies to medicine (n~=~10, 24\%) and economics (n~=~13, 14\%). Other applications were most prominent in political science (n~=~56, 92\%) and economics (n~=~81, 87\%).

\begin{figure}[h]
    \includegraphics[width=\textwidth]{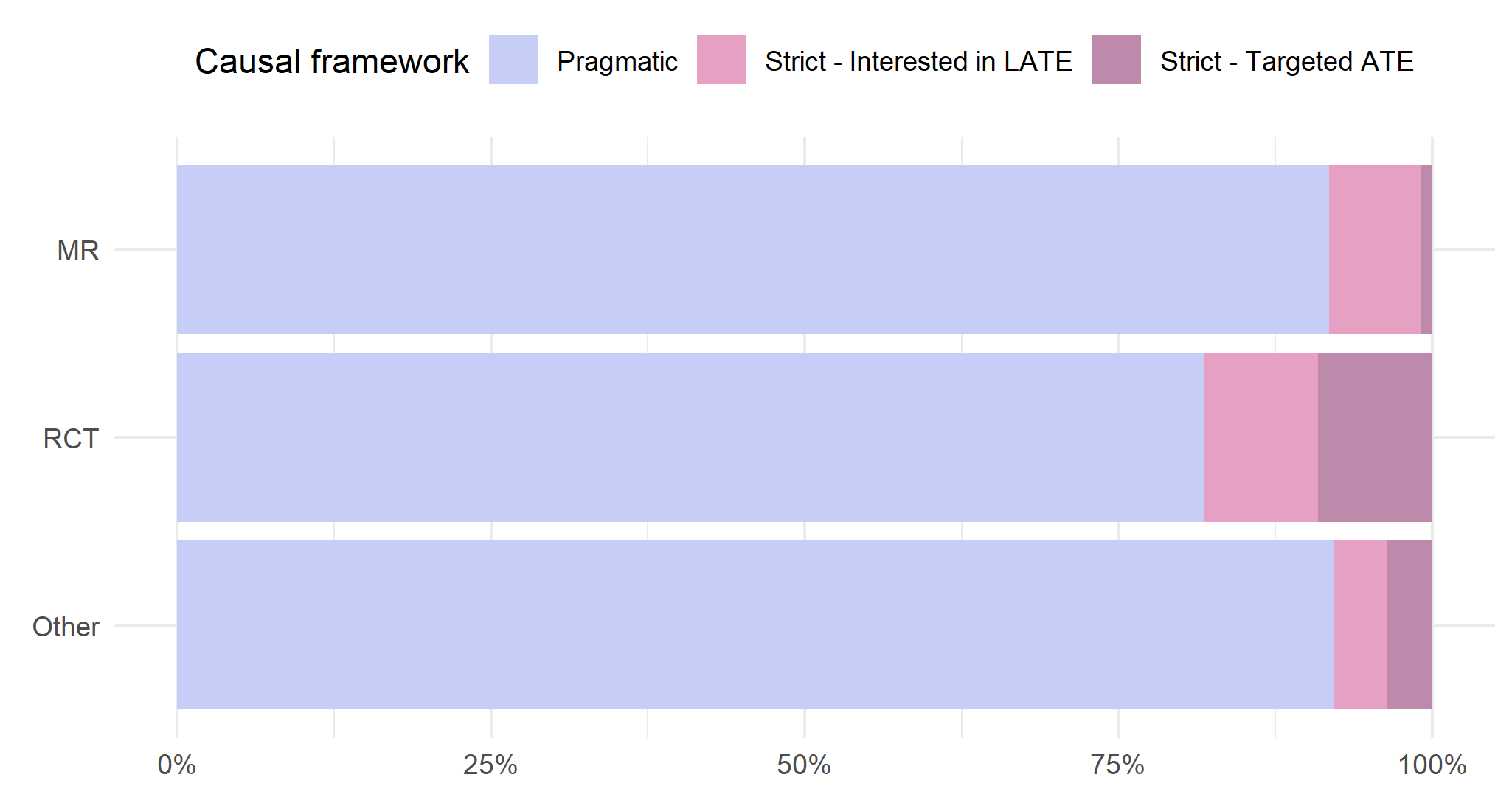}
\caption{Prevalence of causal frameworks by application type.}
\label{fig:results_framework_application_type}
\end{figure}

\begin{figure}[h!]
    \includegraphics[width=\textwidth]{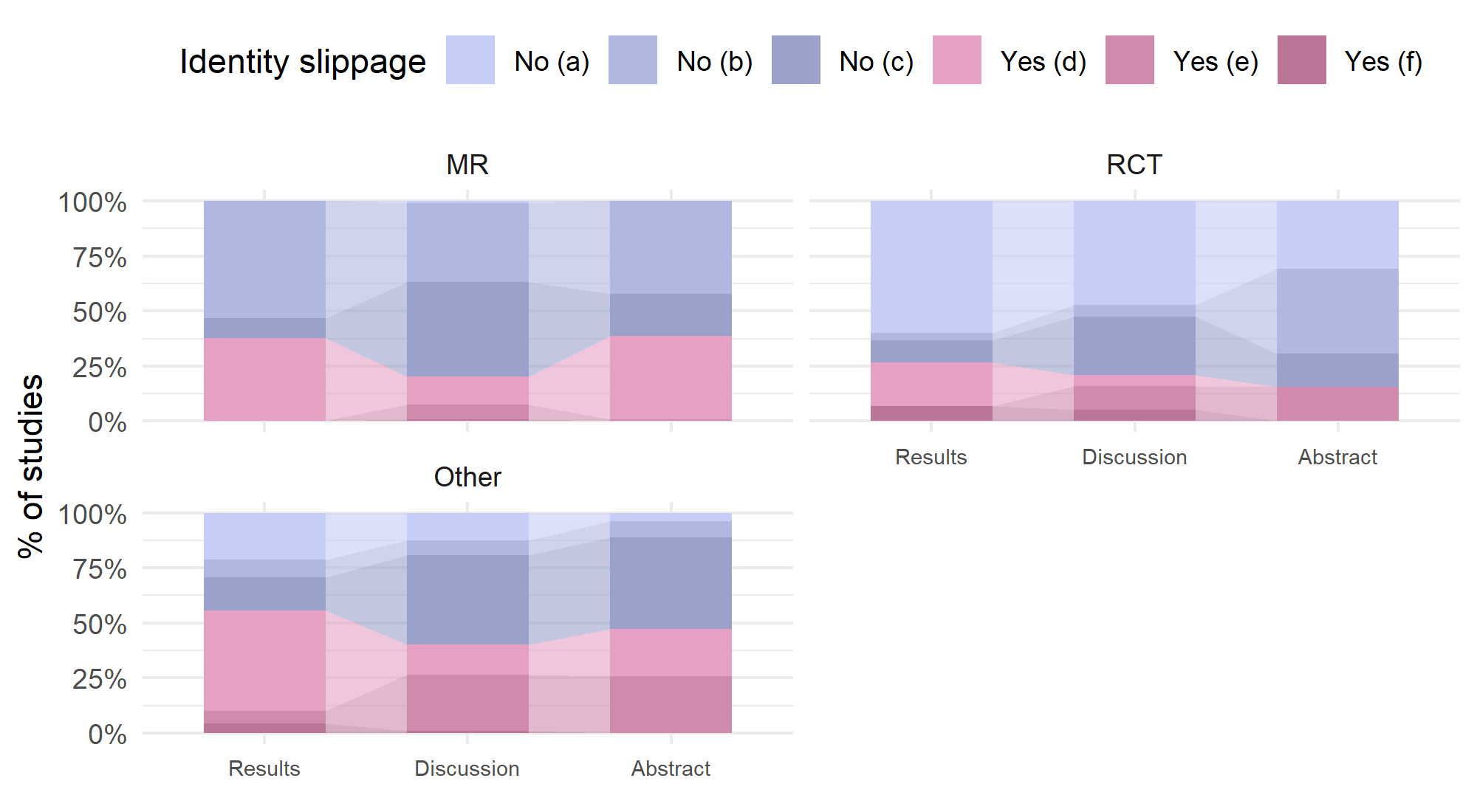}

\caption{Prevalence of identity slippage across article section by application types. Letter in parentheses denotes claim type.}
\label{fig:results_slippage_application_types}
\end{figure}

Figure \ref{fig:results_framework_application_type} illustrates the distribution of \textit{pragmatic} and \textit{strict} studies across application types. Across types of application, \textit{ideal} studies were relatively rare. The highest share of \textit{ideal} studies was observed for RCTs (n~=~6, 18\%), followed by MR (n = 9, 8\%), and other applications (n = 13, 8\%). Figure \ref{fig:results_slippage_application_types} illustrates the prevalence of \textit{identity slippage} across application types.\textit{Identity slippage} occurred most frequently other applications (n~=~113, 71\%), followed by MR applications (n~=~61, 55\%) and RCTs (n~=~9, 30\%). Due to the strong correlation between discipline and application type, it remains unclear whether these differences were driven by discipline or application specific factors. The distribution of claim types also varied considerably. Most notable were the literature-specific differences in how authors referenced the local nature of $\LATE$. In MR applications, all such references were of type (b). In RCTs and other applications, such references were predominantly of type (a).

\subsubsection{Sensitivity analyses}
To test the robustness of our results to alternative specifications of potentially controversial classification criteria, we performed two sensitivity analyses.

In our main analysis, linear regression equations were not considered as homogeneity assumptions that allow for point identification of $\ATE$. Here, we considered linear regression equations as such assumptions. In economics, 66 articles (69\%) contained a linear regression equation. Of these studies, 30 (45\%) contained an explicit reference that $\LATE$ was targeted.\footnote{Indicated by either an explicit specification of the target estimand in the description of the statistical methodology, or at least one claim of type (a) or (b)} For the remaining studies, we reclassified $\targetedEstimand = \ATE$. Compared to the main analysis, more studies were classified as \textit{ideal} (n~=~47, 51\%). According to this alternative specification, $\LATE$ was targeted in 52 studies (56\%), of which 29 (56\%) exhibited at least one occurrence of \textit{identity slippage}. This proportion is slightly lower than in the main analysis (n~=~60, 68\%). Considering the whole sample, 103 studies (33\%) specified a linear regression equation. Of these studies, 47 (46\%) contained an explicit reference that $\LATE$ was targeted. Compared to the main analysis, more studies were classified as \textit{strict} (n~=~84, 27\%). According to this alternative specification, $\LATE$ was targeted in 243 studies (79\%), of which 135 (66\%) exhibited at least one occurrence of identity slippage. Again, this proportion is slightly smaller than in the specification used in the main analysis (n~=~183, 61\%).

For some claims, assignment to a class involved ambiguity. Here, we reclassify claims that we considered to be at the border of whether they implied a presumed identification of $\ATE$. We thus provide a lower bound that represents a specifications where all claims that were at the borderline were reclassified such that $\claimedEstimand = \LATE$, and an upper bound where these claims were reclassified such that $\claimedEstimand = \ATE$. In economics, this leads to a lower bound of 56 studies (60\%) that contain at least once instance of \textit{identity slippage}, and an upper bound 62 studies (70\%). In the whole sample, this lower bound is 173 studies (58\%), and the upper bound is 190 studies (64\%).

\end{document}